\documentclass[aps,prb,10pt,twocolumn,amsmath]{revtex4-1}

\usepackage{graphicx}

\newcommand{\next}{\ensuremath{n_\mathrm{e}}}
\newcommand{\uext}{\ensuremath{u_\mathrm{c}}}
\newcommand{\meV}{\ensuremath{\mathrm{meV}}}
\newcommand{\cmsq}{\ensuremath{\mathrm{cm}^{-2}}}

\newcommand{\eip}{\ensuremath{e^{i\phi}}}
\newcommand{\emip}{\ensuremath{e^{-i\phi}}}

\newcommand{\dvec}[1]{\ensuremath{\boldsymbol{#1}}}
\newcommand{\vk}{\dvec{\mathrm{k}}}

\begin{document}

\title{$\frac{d\mu}{dn}$ in suspended bilayer graphene: the interplay of
disorder and band gap}
\author{D. S. L. Abergel}
\author{H. Min} 
\altaffiliation[Current address: ]{Department of Physics and Astronomy,
Seoul National University, Seoul 151-747, Korea.}
\author{E. H. Hwang}
\author{S. Das Sarma}
\affiliation{Condensed Matter Theory Center, Department of Physics,
University of Maryland, College Park, MD, 20742, USA}
 
\begin{abstract} 
We present an interpretation of recent experimental measurements of 
$\frac{d\mu}{dn}$ in suspended bilayer graphene samples. 
We demonstrate that the data may be quantitatively described by assuming
a spatially varying inter-layer potential asymmetry (which generates a
band gap) induced by local electric fields resulting from charged
impurity disorder in the graphene environment.
We demonstrate that the fluctuations in the inter-layer potential
asymmetry and density vary between different samples, and that the
potential asymmetry fluctuations increase in magnitude as the density
fluctuations increase. 
This indicates that the mechanism causing this effect is likely to be
extrinsic.
We also provide predictions for the optical conductivity and mobility of
suspended bilayer graphene samples with small band gaps in the presence
of disorder.
\end{abstract}

\maketitle

Recently, the fabrication of suspended bilayer graphene (BLG) devices
\cite{martin-prl2010, yacoby-pc} has allowed the direct local
measurement of the thermodynamic quantity $K\equiv\frac{d\mu}{dn}$ using
scanning single electron transistor spectroscopy.
This is an important quantity to study since it is linked directly to
thermodynamic observables such as the thermodynamic density of states,
electronic compressiblility and the quantum capacitance of the
interacting electron liquid in the BLG. 
In samples mounted on a SiO$_2$ substrate, the level of disorder is
generally too high for the intrinsic physics near the charge neutrality
point to be seen because of the existence of strong inhomogeneity
(electron-hole puddles) induced by the disorder\cite{rossi-prl2008}. 
However, suspended samples provide a setting where the effective
strength of the impurities is reduced as reflected, for example, in the
enhanced mobility\cite{bolotin-ssc2008, adam-ssc2008} and it is likely
that the low-density physics is more readily accessible.
In the experiments\cite{martin-prl2010,yacoby-pc}, a region of decreased 
compressiblilty (or equivalently a peak in $K$) is seen near the
charge-neutrality point in zero magnetic field indicating the possible
presence of a small gap in the low energy band structure. 
It is claimed \cite{martin-prl2010} that the size of
the gap which generates this peak is approximately $2\meV$, and that the
magnitude of charge inhomogeneity is of the order of $10^{10}\cmsq$. 
The authors of Ref.~\onlinecite{martin-prl2010} state
that Castro \textit{et al.} \cite{castro-prl2007} show
that this level of in-plane charge inhomogeneity by itself would create
a gap which is a factor of 10 smaller than that observed, ruling out
disorder-induced fluctuations as a mechanism for the generation of this
gap. Instead, they claim\cite{martin-prl2010} that this peak is
evidence for a correlated ground state of the interacting electron
liquid, such as one of those proposed in the recent literature
\cite{nilsson-prb2006, min-prb2008, zhang-prb2010, nandkishore-prb2010,
vafek-prb2010, lemonik-prb2010, nandkishore-prl2010, jung-prb2011,
fan-prl2011}. 
Some of these proposed states may induce a spontaneous many-body charge
transfer instability in the bilayer system which could open a small band
gap at zero density, and the region of decreased compressibility
seen in the experiments may be evidence for the formation
of one of these charge-transfer-instability states \cite{martin-prl2010}.
However, we believe that this interpretation of the experimental results
is problematic because the out-of-plane charge imbalance which would
drive a disorder-induced gap is not the same quantity as the inplane
charge inhomogeneity which Martin \textit{et al.}
use\cite{martin-prl2010} to estimate the gap size.
Therefore it is important to look more closely at the role of
disorder in the suspended bilayer graphene system to investigate in
depth the issue of the possible existence (or absence) of a many-body
quantum phase transition creating a spontaneous band gap.
The goal of the current work is to carry out such a phenomenological
investigation to see whether random charge impurities in the bilayer
environment could produce a signature in the compressibility consistent
with the observations of Refs.~\onlinecite{martin-prl2010,yacoby-pc}.

Transport measurements suggest that there are some residual impurities
in the environment of the suspended graphene, each of which will have an
electric field associated with it, and it is well established that the
electrons in the BLG order themselves into puddles of electrons and
holes in order to screen these fields \cite{dassarma-prb2010}.
Additionally, it was shown recently \cite{rutter-natphys2011} that these
electric fields also cause a spatial variation in the interlayer
potential asymmetry which is highly correlated with the disorder
profile. This is a different effect from the inplane reorganization of
charge, and the two effects may in principle be of quite different
magnitudes, and both are likely to be present in disordered bilayer
graphene. Although this interlayer effect was reported for bilayer
graphene supported on a substrate\cite{rutter-natphys2011}, the general
principle applies to suspended samples as well.

In this article, we revisit the issue of density fluctuations induced by
charged impurity disorder, and show that our calculation of $K$ by
averaging over the density of the electrons in the
puddles\cite{abergel-prb2011} can be applied in two different
phenomenological ways to this situation and gives results for the gap
fluctuations, density fluctuations, and required charge imbalance which
are self-consistent.
By fitting this theory to measurements of $K$ from two different
devices\cite{martin-prl2010,yacoby-pc}, we show that the gap and charge
inhomogeneity are device-dependent, which is, of course, not unexpected
since disorder in different devices is likely to be different leading to
different interlayer potential asymmetry in each sample. 
We also demonstrate that the transport characteristics associated with
our phenomenological disorder fits are reasonable, and predict that in
an optical spectroscopy experiment, the gap will still be obscured by
disorder in these samples\cite{min-prb2011}.
Since our results are internally consistent, we claim that the mechanism
of local density fluctuations induced by charged impurity disorder
cannot be ruled out as a cause of the
observed\cite{martin-prl2010,yacoby-pc} peak in $K$ at low carrier
density.
Our work by no means eliminates the possible occurence of a many-body
BLG instability, it just establishes the qualitative and quantitative
importance of disorder in thinking about the quantitative aspects of the
observed effects, and points out that there is a possible
single-particle extrinsic cause for the observed effective gap, namely,
the inter-layer potential asymmetry induced by the random charged
impurities in the BLG environment. In reality, both interaction and
disorder may be present in a non-perturbative manner, making a
microscopic theoretical analysis quite challenging, particularly since
the graphene landscape becomes inhomogeneous due to the formation of
electron-hole puddles in the presence of the charged impurity disorder.

In order to model the clean BLG system, we consider the continuum limit
of the tight-binding Hamiltonian. In this approach, the wave functions
are described by Bloch functions modulated by a spinor, the elements of
which correspond to the four sites in the BLG unit
cell\cite{abergel-advphys2010}. 
We make this assumption despite the fact that the inhomogeneity in the
disorder potential breaks translational symmetry because we assume that
the changes in the potential landscape are spatially slowly varying
enough that the physics associated with p--n junctions or quantum dots
do not manifest. 
We also point out that this theory has been very successful in describing
capacitance-based compressibility experiments in samples which are much
dirtier than those we currently discuss\cite{abergel-prb2011}.
The band structure is determined by the relative strength of
electron hopping between the four lattice sites.
To simulate the experimental situation as accurately as possible, we go
beyond the standard nearest-neighbor tight binding formalism 
(where $v_F$ is the Fermi velocity of monolayer graphene which comes
from the intra-layer hopping between adjacent atoms and $\gamma_1$ is
the energy associated with the inter-layer dimer bond
which generates the finite effective mass at low energy)
by including the next-nearest neighbor inter-layer hops with energy
$\gamma_4$ and the onsite energies given by the lattice site asymmetry
$\Delta$ and a potential asymmetry $u$ between the two layers. 
This potential asymmetry directly generates a band gap at the K point of
the band structure, and we shall therefore use the terms `inter-layer
potential asymmetry' and `gap' synonymously in this paper.
The Hamiltonian for one valley is then
\begin{equation}
	H = \begin{pmatrix} \frac{u}{2} & 0 & -\hbar v_4 k \emip & 
		\hbar v_F k \emip \\
	0 & -\frac{u}{2} & \hbar v_F k \eip & -\hbar v_4 k \eip \\
	-\hbar v_4 k \eip & \hbar v_F k \emip & -\frac{u}{2} + \Delta &
		\gamma_1 \\
	\hbar v_F k \eip & -\hbar v_4 k \emip & \gamma_1 & 
		\frac{u}{2} + \Delta
	\end{pmatrix}
	\label{eq:Hamiltonian}
\end{equation}
where $v_4 = \sqrt{3}\gamma_4 a/2\hbar$ with $a$ the lattice
constant, and $k$ and $\phi$ are respectively the radial and angular
part of the wave vector measured from the K point. 
The spectrum is valley-symmetric and isotropic. 
The effect of the $\gamma_4$ and $\Delta$ terms is to
introduce an asymmetry between the low-energy conduction and valence
bands, and has previously been used to explain an observed asymmetry in
capacitance measurements \cite{henriksen-prb2010}. Using the results
collated in Ref. \onlinecite{abergel-advphys2010} we choose
representative parameters $\gamma_1=0.4\mathrm{eV}$, $\gamma_4 =
0.15\mathrm{eV}$, $\Delta = 0.018\mathrm{eV}$, $v_F=10^6
\mathrm{ms}^{-1}$ and $a=0.246\mathrm{nm}$.
We describe the band structure and $K$ in the absence of
disorder in the appendix.

The existence of puddles of carriers in graphene is a well-established
phenomenon both in experiment and in theory\cite{dassarma-rmp2011}.
In a previous paper\cite{abergel-prb2011}, we introduced the macroscopic
averaging over density fluctuations and applied this theory to
measurements of the capacitance of BLG devices finding good agreement
with the experimental data. The idea of the theory is that the potential
fluctuations associated with charged impurity disorder generates a
spatially-varying on-site term in the tight-binding Hamiltonian.  This
in turn leads to a local fluctuation in the Fermi energy and hence in
the carrier density.
The statistics of this fluctuation can be encapsulated within a
distribution function $P$. 
The coupling between the tip of the SET and the electron liquid in the
bilayer graphene is capacitative, so we assume that each region with a
given carrier density has a local capacitance associated with it. Then,
the total capacative coupling of the SET to the sample is the parallel
coupling of all of the areas, which equates to the geometrical average
over the region sampled by the SET. Therefore, the areal average
can be replaced by an average over the charge density distribution $P$.
We note that the dimension of the region sampled by the SET in the
experiments of Ref. \onlinecite{martin-prl2010} is $\sim100\mathrm{nm}$
whereas the puddle size seen in Ref. \onlinecite{rutter-natphys2011} is
$\sim 10\mathrm{nm}$ so that the averaging should be reasonable in this
case.
We stress that this philosophy deliberately retains the inhomogeneity of
the system and treats it as the primary manifestation of disorder. This
is a fundamentally different approach from, for example, a diagrammatic
expansion of the electron--impurity interaction which contains within it
an average over realizations of disorder which explicitely restores the
translational symmetry.
We believe that the important disorder physics to be included in the
low-density regime of graphene near the charge neutrality point is the
formation of electron-hole puddles in the system around the
discrete quenched charged impurity centers leading to strong spatial
inhomogeneity modelled by the distribution function $P$ mentioned above
which has been calculated in the literature in a few instances
\cite{rossi-prl2008} and measured in experiment for a bare SiO$_2$
substrate\cite{burson-unpub}.

For density fluctuations associated with puddles of carriers in graphene
systems we write the local density as $n(r) = \next + \tilde{n}(r)$
where $\next$ is the average density induced by external gates and the
fluctuations are encapsulated within $\tilde{n}$. 
We parameterize these fluctuations by their standard deviation 
$\delta n$, and computing the
average of $K$ over the probability distribution for the density 
results in the following expression for the average inverse
compressibility
\begin{equation}
	\bar{K}(\next,\delta n,u) = 
	\int K(n,u) P(n,\next, \delta n) dn.
	\label{eq:Kbar}
\end{equation}
The distribution function $P$ is talken as the following Gaussian
\begin{equation}
	P(x,x_0,\delta x) = \frac{1}{\sqrt{2\pi}\delta x}
	\exp\left[ -\frac{(x-x_0)^2}{2 (\delta x)^2} \right],
	\label{eq:Pdef}
\end{equation}
since there is compelling theoretical\cite{dassarma-prb2010} and
experimental\cite{burson-unpub} evidence that this is approximately the
correct form.
In this theory, the layer asymmetry $u=\uext$ is constant in space (hence
the subscript `$\mathrm{c}$') and is generated by either external
electric fields (such as those from gates) or by an asymmetry of charge
between the two layers of the BLG generated by the intrinsic
electron-electron interactions.
Analytical evaluation of this procedure is not possible since, for finite
$\gamma_4$ and $\Delta$, the eigenvalues of the Hamiltonian
\eqref{eq:Hamiltonian} are the roots of a quartic polynomial. Therefore,
the expression for $\mu$ as a function of $n$, its derivative, and the
integration in Eq.~\eqref{eq:Kbar} must be calculated numerically.

\begin{figure}[tb]
	\centering
	\includegraphics[]{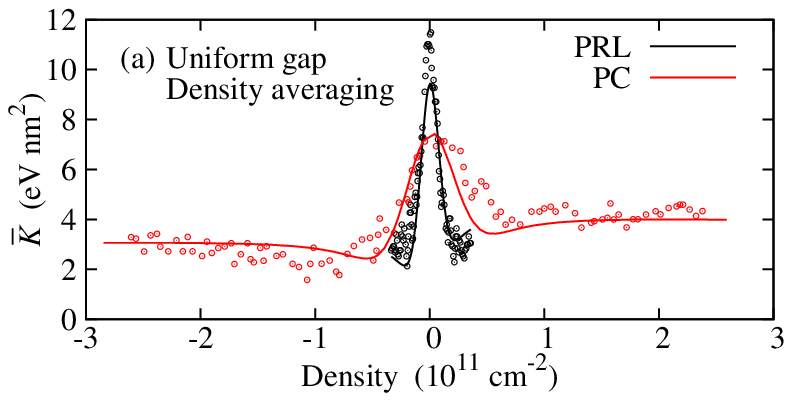}
	\includegraphics[]{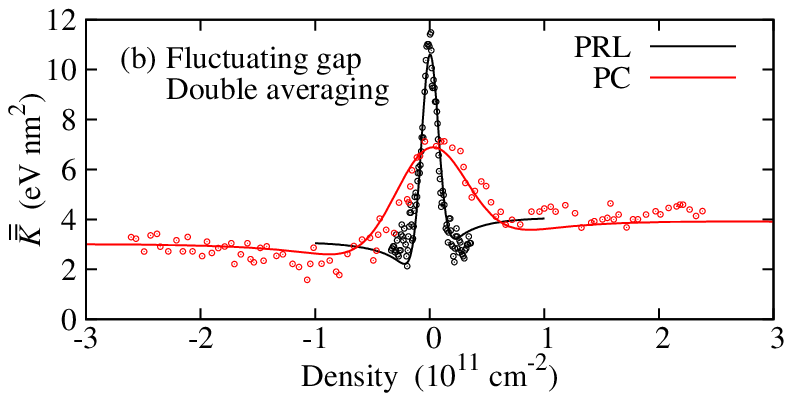}
	\begin{tabular}{|c||c|c||c|c|}
	\hline
	& \multicolumn{2}{|c||}{Density average} & 
	\multicolumn{2}{c|}{Double average} \\ \hline
	& $u_c$ (meV) & $\delta n$ ($10^{10}\mathrm{cm}^{-2}$) & 
	$\delta u$ (meV) & $\delta n$ ($10^{10}\mathrm{cm}^{-2}$) \\ \hline
	PRL & $1.7$ & $0.8$ & $2.2$ & $0.75$ \\ \hline
	PC  & $3.3$ & $2.1$ & $5.2$ & $3.2$ \\ \hline
	\end{tabular}
	\caption{(a) Fits of the density averaged $\bar{K}$ from Eq.
	\eqref{eq:Kbar} to the experimental data. Note that this theory
	includes a spatially uniform band gap $u_c$.
	The black lines are for published data \cite{martin-prl2010}, 
	and the red lines for the unpublished data\cite{yacoby-pc}.
	(b) Fits of the double averaged $\bar{\bar{K}}$ from Eq. 
	\eqref{eq:Kbarbar} with $\uext=0$ to the experimental data. Note
	that in this theory, the band gap $u=\tilde{u}(r)$ fluctuates in
	space, and the distribution of these fluctuations is parameterized
	by $\delta u$.
	The table shows a comparison of the fitting parameters for the two
	samples and two procedures.
	\label{fig:averagefits}}
\end{figure}

Figure \ref{fig:averagefits}(a) shows the results of finding the best
fit between Eq. \eqref{eq:Kbar} and the experimental
data\cite{martin-prl2010,yacoby-pc}. We find that the published
data\cite{martin-prl2010}
(which we shall denote by the label PRL) is quantitatively described by
a constant gap $\uext=1.7\meV$ with density fluctuations
characterized by $\delta n=0.8\times10^{10}\cmsq$, which is very close
to the values claimed by Martin \textit{et al.} in their
paper\cite{martin-prl2010}.
The unpublished data\cite{yacoby-pc} (denoted by PC, as in `private
communication') is described by $\uext=3.3\meV$ and $\delta
n=2.1\times10^{10}\cmsq$. 
Therefore, a different degree of in-plane charge homogeneity and
different size of gap is required to describe the experimental data in
each sample which is perfectly reasonable since we expect the disorder
to have sample-to-sample variations.
Also, the band gap required to fit the data is larger for the more
strongly disordered sample. The opposite would be true if the gap was
the result of an intrinsic interaction-driven effect since in this case,
the gap due to interactions would be universal, but disorder would work
to reduce its size\cite{min-prb2011}.
Therefore, these results, particularly the fact that the more strongly
disordered sample exhibits the larger effective experimental band
gap, indicate that there is some role being played
by extrinsic effects in these experiments.

We now introduce another step in this theory, motivated by recent
observations of spatial fluctuations in the interlayer potential
asymmetry in bilayer graphene \cite{rutter-natphys2011}. 
The interlayer asymmetry was measured in electron puddles and in hole
puddles of a bilayer graphene flake which was mounted on an SiO$_2$
substrate. 
The difference in the asymmetry of approximately $70\meV$ was seen
between the two puddles at zero back gate voltage, and an associated
band gap was seen in the spectrum of Landau levels measured via STM.
Although we expect that the magnitude of this effect will be smaller in
suspended samples, the presence of disorder-induced puddles (as
demonstrated above by our fitting to the appropriate theory) means that
it is sensible to assume that the same disorder may also induce (albeit
smaller in magnitude) local fluctuations in the potential asymmetry just
as it does in graphene on substrates in
Ref.~\onlinecite{rutter-natphys2011}. 
In this case, the potential asymmetry becomes a function of position
such that $u \equiv u(r) = \uext + \tilde{u}(r)$.
The fluctuations are contained within $\tilde u$, and we
assume that they can be described by the distribution $P$ with standard
deviation $\delta u$ (just as the density fluctuations can) because the
same disorder is generating both types of fluctuation. 
Note that it is the in-plane component of
the electric field associated with disorder which causes the formation
of puddles, while the spatially varying inter-layer potential asymmetry
is driven by the out-of-plane component of the electric field.
Therefore, while we expect
the two types of fluctuation to be correlated, it is not obvious exactly
what relationship should exist between them.
However, on a phenomenological level, we can treat $\delta n$ and the
potential asymmetry fluctuations parameterized by $\delta u$ as
independent parameters.
We can therefore perform an average over the gap fluctuations in the
same way we previously did for the density. The spatially varying part
of the band gap $\tilde u$ is described by $\delta u$, and the
Gaussian is centered around \uext.
The spatial average then corresponds to the following average over the
distribution function:
\begin{equation}
	\bar{\bar{K}}(\next,\delta n,\uext,\delta u) = 
	\int \bar{K}(\next,\delta n,u) P(u,\uext,\delta u) du.
	\label{eq:Kbarbar}
\end{equation}
We call this procedure `double averaging'.
In order to assess if the observed gap might be entirely due to
disorder-induced fluctuations, we fit this expression with $\uext=0$ to
the experimental data. 
Therefore the fitting parameters are now $\delta
u$ and $\delta n$, so we have two parameters as was the case for the
density averaging procedure.
The results are displayed in Fig.  \ref{fig:averagefits}(b). 
We find that the PRL data is fitted by $\delta u = 2.2\meV$ and $\delta
n = 0.75\times 10^{10}\cmsq$ and the PC data by
$\delta u = 5.2\meV$ and $\delta n = 3.2\times10^{10}\cmsq$. 
The fit of the double-averaged theory is actually slightly better than
that which is averaged only over density (especially in the region of
the peak), but both averaging procedures give results which are
in reasonable agreement with the experiment and therefore it is not
possible to distinguish between a spatially fluctuating disorder-induced 
gap and a uniform intrinsic gap in the current data\cite{gapexp}.
Therefore, our work convincingly demonstrates they key importance of
disorder, particularly the density and potential fluctuations associated
with the
inhomogeneous puddles, in determining the experimental compressibility
measurements of Refs.~\onlinecite{martin-prl2010,yacoby-pc}.

\begin{figure}[tb]
	\centering
	\includegraphics[]{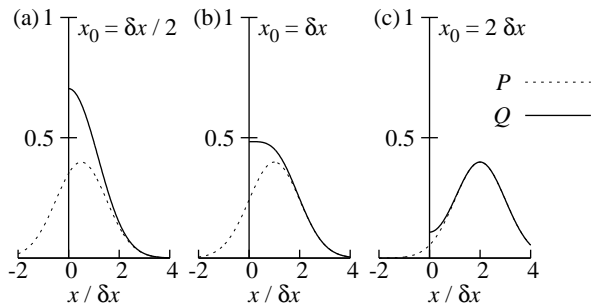}
	\caption{Plots of distributions $P$ and $Q$ for (a) $x_0<\delta x$, 
	(b) $x_0 = \delta x$, and (c) $x_0 > \delta x$. In all cases, we
	have $\delta x=1$.
	\label{fig:gaussians}}
\end{figure}

Using the parameters from these fits, we can make the following
observation. In principle, if there was truly no disorder then both
$\delta n$ and $\delta u$ should be zero. However, if we take the two
pairs extracted from the double averaging fitting procedure and assume a
linear extrapolation to $\delta n=0$, we find a residual $\delta u
\approx 1.3\meV$. 
This may indicate the presence of an intrinsic (possibly, many-body) gap
not generated by the fluctuations, so we examine that issue now.
The experimental measurements of $K$ in the PRL data suggest that the
size of the uniform gap, if it exists, is approximately $2\meV$.
Since this is of the same order as the gap predicted by extrapolating
to the clean limit, we have run fits of our double-averaged theory with
a constant gap of this size to the experimental data. 
We find that there is no improvement in the fitting for any value of
$\delta n$ or $\delta u$.
To explain this, we need to analyze the distribution function for the
gap fluctuations. We know that $K$ is an even function of the gap size
$u$, and therefore the absolute value $|u|$ is the key for determining
the distribution of $K$. Therefore, although we average over the
Gaussian $P$ in Eq. \eqref{eq:Kbarbar}, the effective distribution of
$K$ is non-Gaussian, and given by
\begin{equation}
	Q(x,x_0,\delta x) = 
		P(x,x_0,\delta x) + P(-x,x_0,\delta x),
	\label{eq:Pux}
\end{equation}
where $x>0$ and $P(x,x_0,\delta_x)$ is defined in Eq. \eqref{eq:Pdef}.
Figure \ref{fig:gaussians} shows the distribution $Q$ 
for the cases $x_0 < \delta x$, $x_0 = \delta x$, and $x_0 > \delta x$.
We see that in all but the last case, the distribution is dominated by
small values of $x$ because of the significant contribution from the
$P(-x)$ term, and it is only when $\delta x > x_0$ that the Gaussian
shape is revealed in $Q$. 
In the two sets of data which we have studied, we see that $\uext$ found
from the single averaging procedure and the $\delta u$ from the double
averaging obey the relation $\uext<\delta u$. 
This suggests that the disorder is too strong to determine if a
spontaneous gap exists in the absence of disorder.
In order to measure the spontaneous gap, samples with levels of disorder
such that $\delta u < \uext$ are needed.
This necessitates the development of samples with much higher quality
(i.e. less disorder), perhaps mobilities which are a factor of five or so
larger, to decisively establish the existence of a spontaneous many-body
BLG gap. In short, the disorder-induced gap needs to be a factor of 2
or more smaller than the many-body gap for an unambiguous conclusion on
this issue.

\begin{figure}[tb]
	\centering
	\includegraphics[]{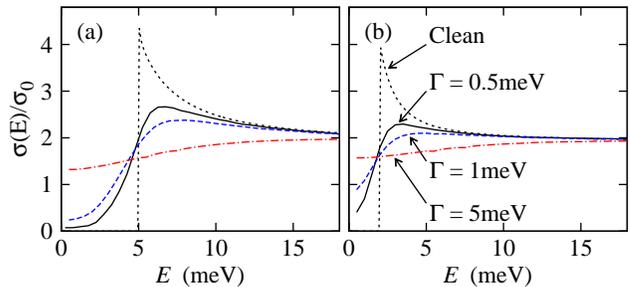}
	\caption{Optical conductivity for (a) $\uext=5\meV$ and (b)
	$\uext=2\meV$ for various degrees of disorder broadening. The color
	of the lines is the same in both panels. Dashed lines indicate the
	clean case.
	\label{fig:optcond}}
\end{figure}

The gap size may also be extracted from measurements of the optical
conductivity via absorption or reflection experiments
\cite{zhang-nature2009,mak-prl2009,kuzmenko-prb2009,li-prl2009}. 
To demonstrate that this may be an unreliable process in the case of the
small band gaps (of the order of a few meVs) we are discussing in this
context, we plot in Fig. \ref{fig:optcond} the optical conductivity
derived from the Kubo formula with $u=5\meV$ in panel (a) and with
$u=2\meV$ in panel (b). 
By way of comparison, in the linear (high density) part of the
hyperbolic spectrum of bilayer graphene, the density fluctuations
$\delta n$ correspond to fluctuations in the Fermi level of
approximately $\hbar v_F \sqrt{\pi \delta n}$, and with $\delta
n=3\times 10^{10}\cmsq$, this gives an energy scale of approximately
$20\meV$. 
When the Fermi energy is in the sombrero (low density) region with
$u=5\meV$, the flatness of the bands means that this density fluctuation
corresponds to a fluctuation of less than 1\meV.
We include the disorder through a constant broadening of the bands
parameterized by the energy $\Gamma$ which appears in the Green's
function as
\begin{equation}
	G_{\vk\lambda}(E) = \frac{1}{ E - E_{\vk\lambda} 
	+ i\Gamma}. \label{eq:Gdef}
\end{equation}
The optical conductivity is calculated from this Green's function from
the following expression\cite{abergel-arXiv}:
\begin{multline}
	\sigma(E) = 2g_s g_v \frac{e^2}{h} \sum_{\lambda,\lambda'}
	\int \frac{k'\,dk'}{2\pi} \int_{E_F-E}^{E_F} \frac{dE'}{E} \\
	\times M^2_{\lambda\lambda'}(k') \mathrm{Im}\,G_{\vk'\lambda}(E')
	\mathrm{Im}\,G_{\vk'\lambda'}(E'+E)
\end{multline}
where $g_s$ and $g_v$ are the spin and valley degeneracies, and
\begin{equation}
	M^2_{\lambda\lambda'}(k) = \int_0^{2\pi} \frac{d\phi}{2\pi}
	\left| \left\langle \lambda, k, 0 | \hbar \hat{v}_x |
	\lambda',k,\phi\right\rangle \right|^2.
\end{equation}
Figure \ref{fig:optcond} shows that for $\Gamma=1\meV$, the peak due to
the onset of the intraband conductivity is already significantly
blurred. 
By the time $\Gamma>\uext$, the peak is completely smeared, as shown by
the $\Gamma=5\meV$ line. 
We mention in this context that the same conclusion can also be reached
by considering an inhomogeneous broadening effect rather than a
homogeneous broadening as used in Eq.~\eqref{eq:Gdef} -- for example, we
can introduce a density fluctuation leading to an uncertainty in the
Fermi energy by 1meV, leading to exactly the same conclusion that the
optical gap measurement would be unable to discern the small band gap in
the presence of this inhomogeneous broadening effect.



\begin{figure}[tb]
	\centering
	\includegraphics[width=\columnwidth]{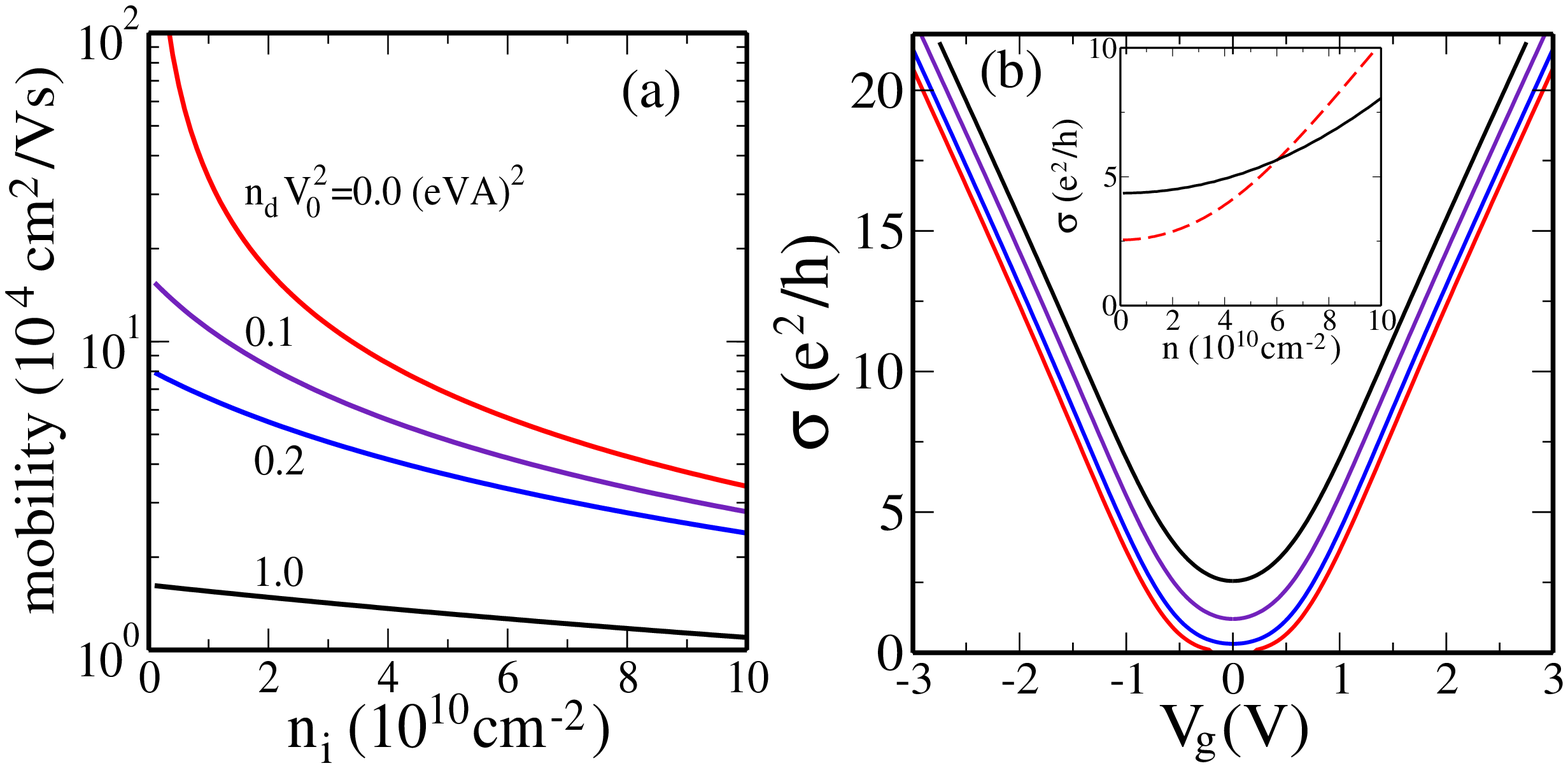}
	\caption{(a) Calculated mobility of suspended bilayer graphene as a
	function of impurity density for different strengths of the 
	short-ranged impurities. 
	(b) Conductivity calculated within effective medium theory as a
	function of gate voltage $V_g$ for a charged impurity density,
	$n_i=2\times 10^{10}\cmsq$, and a fixed short range potential
	$n_dV_0^2=0.3$ (eV\AA)$^2$. Different curves correspond to the band
	gap $E_g = 0$, 0.15, 0.3, 0.33 eV (from top to bottom).  Inset shows
	the calculated conductivities as a function of carrier density for
	two different charged impurity densities $n_i=2\times 10^{10} \cmsq$
	(dashed line) and $n_i=4\times 10^{10} \cmsq$ (solid line) with
	zero band gap. 
	\label{fig:sig}}
\end{figure}

We also predict the transport properties of suspended bilayer graphene
with disorder so that independent characterization of the samples in
which $K$ was measured can be done and compared with our theory
incorporating inhomogeneous puddle effects of density fluctuations. 
We use the highly successful Boltzmann-RPA formalism, which has been
used to study the transport in monolayer graphene and BLG systems
\cite{dassarma-rmp2011}.
This theory incorporates both long-range Coulomb impurities and
short-range scatterers.
In Fig.~\ref{fig:sig}(a) we show the mobility of suspended
BLG as a function of charged impurity density $n_i$, calculated at a
carrier density $n=10^{11} \mathrm{cm}^{-2}$ for different short-ranged
disorder potentials, $n_d V_0^2$, where $n_d$ is the density of
short-ranged impurities and $V_0$ is the strength of the potential.
If we assume a charged impurity density $n_i \approx \delta n$,
then we have $\mu \sim 4-5\times 10^4$ cm$^2$/Vs with a fixed short
range potential $n_dV_0^2=0.3$ (eV\AA)$^2$. 
However, we note that even though the carrier density fluctuation is
related to the impurity density, the theoretical relation between two
densities is not yet known. In Fig.~\ref{fig:sig}(b) we show the
conductivity calculated within effective medium
theory\cite{dassarma-prb2010} as 
a function of gate voltage $V_g$ in the presence of an energy band gap.
We assume that the energy dispersion of carriers is given by $E_{k\nu}
= \nu \hbar^2 k^2/(2m)+ \nu E_g/2$, where $\nu =\pm 1$ indicates the
electron ($\nu=1$) or hole ($\nu=-1$) band, $m$ is the
effective mass, and $E_g$ is the band
gap. In this figure a charged impurity density $n_i=2\times
10^{10}\cmsq$ and a fixed short range potential $n_dV_0^2=0.3$
(eV\AA)$^2$ are used. Our calculation of Fig.~\ref{fig:sig}(b) shows
that due to the electron hole puddles the transport gap (i.e. the
region of zero conductivity) does not 
occur even though there is a band gap (top three curves in the
figure). Only large enough energy gaps (the lowest curve)
reflect the transport gap. However, our calculation also shows that for
small density fluctuations (i.e. shallow 
electron-hole puddles) the transport gap can be induced by the smaller
energy gap.
In the inset to Fig.~\ref{fig:sig}(b) the conductivity of suspended BLG
calculated within the effective medium theory \cite{dassarma-rmp2011} is
shown for two different charged impurity densities, $n_i=2$ and 4$\times
10^{10}\cmsq$ for a fixed short range potential $n_dV_0^2=0.3$
(eV\AA)$^2$. Our calculation shows the minimum conductivity at the
charge neutral point is higher for a low mobility (more strongly
disordered) sample. 


In conclusion, we have demonstrated that even though suspended graphene
samples are known to be significantly less disordered than those mounted
on substrates, the effect of disorder in recent measurements of
$\frac{d\mu}{dn}$ cannot be ruled out. Using phenomenological averaging
procedures, which specifically incorporate the effects of strong
inhomogeneous broadening arising from electron-hole puddle induced
spatial density and potential fluctuations, we have demonstrated that
current experimental data cannot distinguish between an intrinsic,
spatially uniform band gap generated by
electron-electron interactions, and a spatially fluctuating band gap
induced by disorder. 
Specifically, these disorder-induced fluctuations can account for the
observed data without invoking the presence of an intrinsic
gap\cite{gapexp}.
In order to be confident about the origin of the band gap, higher
quality samples where $\uext > \delta u$ must be fabricated.
One possibility is to carry out measurements on BLG samples fabricated
on BN substrates which typically have much lower charged impurity
disorder \cite{dassarma-prb2011, dean-natnano2010}.

The theory we develop involves exactly one uncontrolled approximation,
assuming the density fluctuations and the fluctuations in $u$ to be
uncorrelated.
Going beyond this approximation would require knowledge about the
details of the local electric fields producing the fluctuations in $n$
and $u$ which is currently unavailable experimentally.
It would be straightforward to include the correlation in our theory if
microscopic information about the correlator between fluctuations in $n$
and $u$ is available from experiments or some other microscopic
calculations.
In the absence of such information we assumed them to be independent.

We also emphasize that the theory presented in this article is
phenomenological and is a first step toward the understanding of the
role of disorder in the experiments of Refs.~\onlinecite{yacoby-pc,
martin-prl2010}. What we have
achieved in this paper is a demonstration that spatial density and
potential fluctuations are capable of producing a spatially fluctuating
gap consistent with the experimental data.  Many questions of details
could be raised with respect to our phenomenological procedure, for
example, the averaging procedure we employ is certainly not unique, but
in the presence of real spatial fluctuations which explicitly break the
translational invariance in graphene near the Dirac point, our procedure
is physically motivated as it averages precisely over the quantities
which fluctuate according to some distribution function which has been
calculated in the literature \cite{dassarma-rmp2011, rossi-prl2008,
dassarma-prb2010, rossi-arXiv}.  
A better (purely numerical) method to tackle the problem would
be to use the techniques developed in Refs.~\onlinecite{rossi-prl2008,
dassarma-prb2010, rossi-arXiv} to explicitly and self-consistently
calculate the real-space electronic structure in the presence of the
electron-hole puddles generated by the random charge impurities, and
then to use this
real-space electronic structure to directly numerically calculate the
compressibility.  Such a calculation is obviously numerically
prohibitively difficult, and our work is a short-cut (albeit a rather
simple one) in carrying out such a numerical procedure.  Since the
Gaussian probability distribution
function we use is numerically well-verified in
Refs.~\onlinecite{dassarma-rmp2011, rossi-arXiv, rossi-prl2008,
dassarma-prb2010}
through self-consistent calculations, we believe that our results,
although phenomenological, should be qualitatively and quantitatively
valid.  The eventual theory involving a microscopic calculation of the
compressibility in the presence of long-range Coulomb disorder and
electron-electron interaction including the realistic band structure of
bilayer graphene is very far in the future since no one knows how to
approach such a non-perturbative problem even at a formal level.  

\begin{acknowledgments}
We acknowledge support from US-ONR and NRI-SWAN and
thank Amir Yacoby for sharing unpublished data with us.
\end{acknowledgments}

\appendix
\section{Clean bilayer graphene}
\begin{figure}
	\centering
	\includegraphics[]{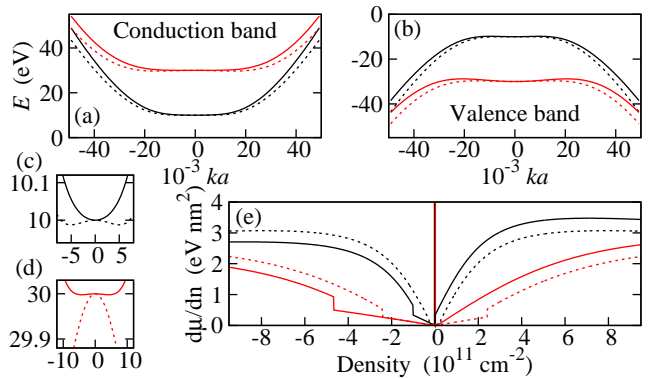}
	\caption{(a,b) The band structure of clean bilayer graphene with a band
	gap. The dashed lines correspond to the electron-hole symmetric band
	structure found with $\gamma_4 = \Delta = 0$ [see
	Eq.~\ref{eq:symenergy}] and the solid line to the full band
	structure with $\gamma_4 = 150\meV$, $\Delta =
	18\meV$. Black lines are for $u=20\meV$ and red lines for
	$u=60\meV$.
	(c) The bottom of the conduction band for $u=20\meV$ showing the 
	absence of sombrero structure for finite $\gamma_4,\Delta$.
	(d) The bottom of the conduction band for $u=60\meV$ showing the
	restoration of the sombrero structure for finite $\gamma_4,\Delta$.
	(e) $\frac{d\mu}{dn}$ for clean bilayer graphene.  The lines
	correspond to the same as in (a).
	\label{fig:cleanbs}}
\end{figure}

The band structure associated with the Hamiltonian presented in
Eq.~\eqref{eq:Hamiltonian} is shown in Fig.~\ref{fig:cleanbs}.
In the symmetric case with $\gamma_4 = \Delta = 0$, the low energy bands
have the dispersion
\begin{equation}
	E_\pm = \pm \sqrt{ \tfrac{\gamma_1^2}{2} + \tfrac{u^2}{4} + \kappa^2
	- \sqrt{ \tfrac{\gamma_1^4}{4} + \kappa^2( \gamma_1^2 +u^2 ) } }
	\label{eq:symenergy}
\end{equation}
where $\kappa = \hbar v_F k$. Writing this as a function of density and
differentiating yields the following analytical expression for
$\frac{d\mu}{dn}$ in this limit in the conduction band:
\begin{equation}
	K = \begin{cases} 
	\frac{\hbar^2 v_F^2 \pi}{2} 
		\frac{\lambda}{\sqrt{u^2+\gamma_1^2}}
		\frac{1}{\sqrt{\lambda^2 + \gamma_1^2 u^2}}
		& \lambda < u^2, \\
	\frac{\hbar^2 v_F^2 \pi}{2} 
		\frac{1 - (u^2+\gamma_1^2)/(2\alpha)}
		{ \sqrt{ \lambda + \frac{u^2}{4} + \frac{\gamma_1^2}{2} -
		\alpha}}
		& \lambda \geq u^2,
	\end{cases}
\end{equation}
where $\lambda = \hbar^2 v_F^2 \pi n$ and $\alpha = \sqrt{
	\lambda(u^2+\gamma_1^2) + \frac{\gamma_1^4}{4}}$.
In contrast, when $\gamma_4$ and $\Delta$ are included in the analysis,
analytical solutions of the eigenvalue problem are not possible. The
energy spectrum is found as the roots of the following quartic
polynomial equation for $E$:
\begin{multline}
	E^4 - 2\Delta E^3 + \left[ \Delta^2 - \gamma_1^2 - \frac{u^2}{2}
	-2\kappa^2\left( 1 + \chi^2 \right) \right] E^2 \\
	+ \left[ \frac{\Delta u^2}{2} + 2\kappa^2\left( \Delta +
	\chi^2\Delta + 2 \chi \gamma_1 \right)
	\right] E \\
	+ \frac{u^2}{4}\left( \gamma_1^2 - \Delta^2 \right) + 
	\left[ \frac{u^2}{4} - \kappa^2\left( 1 - \chi^2 \right) \right]^2 = 0
	\label{eq:g4poly}
\end{multline}
where $\chi = v_4/v_F$. 
The Fermi surface may be ring-shaped for finite values of $u$ so that
the wave vectors corresponding to the inner and outer Fermi surfaces
$k_\pm$ are given by the equation $\pi n = k_+^2 - k_-^2$
with the constraint that $k_+$ and $k_-$ must give the same
energy for the appropriate band when substituted into
Eq.~\eqref{eq:g4poly}. 
When the Fermi energy is above the sombrero region, $k_-=0$. The
extracted value of $k_+$ can be substituted into Eq.~\eqref{eq:g4poly}
to find $\mu$. The differentiation to obtain $K$ can the
be carried out numerically.
Evaluations of these equations are shown in Fig.~\ref{fig:cleanbs} for the
band structure and $K$ for two different values of the
band gap, and with and without the tight-binding parameters $\gamma_4$
and $\Delta$. 
The step feature in $K$ is a result of the change in
topology of the Fermi surface from a ring to a disc as the Fermi energy
leaves the sombrero region.  For small values of $u$, the conduction
band does not show the sombrero shape when $\gamma_4,\Delta$ are finite
and this step is absent.
The spike at the origin is a $\delta$-function resulting from the
discontinuity of the Fermi energy at zero density as it jumps from the
valence band to the conduction band.

\end{document}